\documentclass[aps,preprint,showpacs,showkeys,nofootinbib]{revtex4-1}

\usepackage{amsmath}
\usepackage{amsfonts,amssymb}
\usepackage{multirow}
\usepackage{graphicx,booktabs,rotating}
\usepackage{float}
\usepackage{appendix}
\usepackage{color}
\usepackage{url}
\usepackage{subfigure}
\usepackage{footnote}

\def\beq{\begin{equation}}
\def\eeq{\end{equation}}
\def\bea{\begin{eqnarray}}
\def\eea{\end{eqnarray}}

\def\nl{\nonumber\\}

\def\chic1{\chi_{c1}}

\newcommand{\gsim}{\lower.7ex\hbox{$
\;\stackrel{\textstyle>}{\sim}\;$}}
\newcommand{\lsim}{\lower.7ex\hbox{$
\;\stackrel{\textstyle<}{\sim}\;$}}

\newcommand{\eod}{\end{document}}

\newcommand{\DDs}{D_{(s)}}

\RequirePackage{xspace} \allowdisplaybreaks

\begin{document}

\title{Branching fractions of semileptonic $D$ and $D_s$ decays\\ from the covariant light-front quark model}
\author{Hai-Yang Cheng}
\author{Xian-Wei Kang}
\affiliation{Institute of Physics, Academia Sinica, Taipei, Taiwan
115}

\begin{abstract}
 Based on the predictions of the relevant form factors
 from the covariant light-front quark model, we show the branching fractions for
 the $D (D_s) \to (P,\,S,\,V,\,A)\,\ell\nu_\ell$ ($\ell=e$ or $\mu$)
 decays, where $P$ denotes the pseudoscalar meson,
 $S$ the scalar meson with a mass above 1 GeV, $V$ the vector meson
 and $A$ the axial-vector one. Comparison with the available
 experimental results are made, and we find an excellent agreement. The
 predictions for other decay modes can be
 tested in a charm factory, e.g., the BESIII detector.
 The future measurements will definitely further enrich our knowledge on the hadronic
 transition form factors as well as the inner structure of the even-parity mesons ($S$ and $A$).
\end{abstract}¡¡

\maketitle

%%%%%%%%%%%%%%%%%%%%%%%
\section{Introduction}
\label{Sec:Introduction}
%%%%%%%%%%%%%%%%%%%%
The Cabibbo-Kobayashi-Maskawa (CKM) matrix describing the quark
flavor mixing \cite{CKM} has been a key skeleton of the Standard Model
(SM). Precise determination of its matrix elements is one of the central
tasks for both theoretical and experimental colleagues all along.
Any deviation from the unitarity relation is believed to be an
exciting signal of New Physics (NP). As it is known,  the
semileptonic decay of heavy flavor meson plays an important role in
extracting CKM elements, e.g., the $V_{cd}$ from $c\to d$ decay and
$V_{cs}$ from $c\to s$ decay. The extraction of $V_{cq}$ ($q=d$ or $s$) needs some sophisticated
knowledge of the form factors relevant for the decay process.

We first briefly introduce the form factors to be used in this work.
Among the various models, we will concentrate on the description of
form factors from the covariant light-front quark model (CLFQM). In
1949, Dirac proposed three different forms fulfilling the special
theory of the relativity and the Hamiltonian formulation of the
dynamics \cite{Dirac}: instant form, point form and light-front
form. The light-front form ($x^+=x^0+x^3=0$) has the advantages that
there are only three Hamiltonians from the ten fundamental
quantities in the Poincar\'{e} group and that the square root is
absent in the Hamiltonians such that one can avoid the
negative-energy states. The quark model expressed in the light-front
form constitutes the so-called light-front quark model, which has
been extensively developed to treat the electroweak (radiative and
semileptonic) decays of the mesons in the early 1990s
\cite{Jaus1990,Jaus1991}. In such  theory, one can first draw the
Feynmann diagram and then write down the amplitude. Meson is a
bound state of its quark component $q\bar q$. The vertex function
between the meson and its $q\bar q$ is obtained from the wave
function composed of the momentum distribution of the constituent
quarks in the meson and the spin part. The latter involves the
Melosh-type rotation from the conventional spin state (or the
so-called instant form of the spin state by Dirac) to the one in the
light-front form. In fact, the occurrence of the light-front form
can be also easily seen from the infinite-momentum frame
\cite{Weber}. The quark internal line is just given by the fermion
propagator. Taking the plus component of the corresponding current
matrix element will give the final result for the form factor. Note
that in such a conventional light-front quark model, the internal
quarks are on their mass shells, and the zero-mode effect is missed
which renders the theory non-covariant. Considering these defects,
the covariant light-front quark model (CLFQM) \cite{Jaus1999} was
later proposed (see also more works on this aspect in
\cite{ChoiJi1,ChoiJi2,ChoiJi3} and a very recent work
\cite{Chen:2017vgi}). Following the lines of Ref.~\cite{Jaus1999},
Cheng, Chua and Hwang have systematically studied the decay
constants and form factors for the $S$- and $P$-wave mesons in 2003
\cite{ChengChua2003CLF}, while an update was done in
Ref.~\cite{Verma}. In the latter reference, the author applied the
available experimental information and the lattice results for the
decay constants to constrain (part of) the parameters $\beta$ in the
wave functions, and also incorporated the $D_s$ and $B_s$ decays.
However, only the relevant form factors are presented in
Refs.~\cite{ChengChua2003CLF,Verma}. In this work we shall further
provide the branching fractions which are the {\it true observables}
in experiment such that we can make a {\it direct} comparison
between theory and experiment. Moreover, the large statistics
accumulated by the BESIII is capable of carrying out such a task.

We discuss below how we understand the main usage of the
branching fractions predicted in this work. Our considerations are
as follows:
\begin{itemize}
\item $|V_{cd}|$ and $|V_{cs}|$ are well-determined
quantities, which are also used as input for calculating the
branching fractions.
\item We will see below that the
comparison of the theoretical predictions with experimental
measurements for $P$ and $V$ mesons leads to an excellent agreement. This demonstrates that the CLFQM works very well.
\item To calculate the branching fractions, we have considered the underlying structure
of the final-state meson, e.g., the mixing angles for axial-vector
$f_1$, $h_1$ and $K_1$ mesons. Especially, the scalar isosinglet
$f_0$ states above 1 GeV are considered as the mixture of $q\bar q$
and the glueball state $G$. The confrontation of our theoretical
predictions with experiment will {\it help pin down the issue of the
underlying structures of these mesons}. Emphasis will be put on the
scalar and axial-vector mesons, which are less understood compared
to the pseudoscalar and vector octet ones.
\item The three-body semileptonic decay provides a clean environment
for the study of the weak hadronic transition as well as the
underlying structure of the involved mesons due to the absence of
the final-state interactions (FSIs) between hadrons \footnote{FSIs
in the four-body semileptonic decay mode, e.g. $D\to \pi\pi l\bar
\nu_l$, can be carefully explored following the line sketched in
Ref.~\cite{KangBl4}}.
\end{itemize}

Experimentally, such a goal of testing the inner structure of the axial-vector
 and scalar mesons is doable due to the existing large
statistics of $D$ and $D_s$ mesons. We need to point out that
$1.8\times10^7$ $D^0\bar D^0$ and $1.4\times10^7$ $D^+D^-$ (at
$\psi(3770)$ peak), $2.0\times 10^7$ $D_s^+D_s^-$ pairs (at the center
of mass of 4.18 GeV) will be accumulated per year according to the
design plan of BESIII \cite{BES1,BES2,Front}. For a super tau-charm
factory, the luminosity will be further enhanced by 100 times
\cite{super1,super2}.

The outline of this work is as follows. For completeness, we show in
Sec.~\ref{Sec:FF} the formula for decay rates in details. Then in Sec.~\ref{Sec:results} the results for
the branching fractions are summarized in Table~\ref{tab:BR} for the
electron mode and \ref{tab:BRmu} for the muon mode. The discussions
are also presented there. Sec. IV comes to our conclusions.
%The experimental colleagues working in the field of the charm physics will make a measurement to test our predictions.

%%%%%%%%%%%%%%%%%%%%%%%%%%
\section{Form factors, decay rates}
\label{Sec:FF}
%%%%%%%%%%%%%%%%%%%%%%
We will follow the Bauer-Stech-Wirbel (BSW) model \cite{BSW} for the
convention of form factors and  its extension to scalar and
axial-vector mesons \cite{ChengChua2003CLF}. The pseudoscalar,
scalar, vector, and axial-vector mesons are denoted by
$P,\,S,\,V$ and $A$,  respectively. The form factors are given by
\footnote{Some early studies on the flavor-symmetry breaking of
$D\to P$ form factors can be found in e.g., Ref.~\cite{Khlopov}.}
\begin{eqnarray}
\langle P(p'')|V_\mu|\DDs(p')\rangle
&=&\left(p_\mu-\frac{m_{\DDs}^2-m_P^2}{q^2}q_\mu\right)F_1^{\DDs\to
P}(q^2)\nl && +\frac{m_{\DDs}^2-m_P^2}{q^2}q_\mu F_0^{\DDs\to
P}(q^2)
\end{eqnarray}
for the transition of
$D_{(s)} \to P$, and
\begin{eqnarray}\label{eq:FFDS}
\langle S(p'')|A_\mu|\DDs(p')\rangle
&=&-i\Biggl[\left(p_\mu-\frac{m_{\DDs}^2-m_S^2}{q^2}q_\mu\right)F_1^{\DDs\to
S}(q^2)\nl&&+\frac{m_{\DDs}^2-m_S^2}{q^2}q_\mu F_0^{\DDs\to
S}(q^2)\Biggr].
\end{eqnarray}
for the $\DDs \to S$ transition. In above equations, $V_\mu$ and $A_\mu$ are the corresponding vector and axial-vector
currents dominating the weak decay. The momenta $p$ and $q$ are
defined as $p=p'+p''$ and $q=p'-p''$, where $p' (p'')$ is the
four-momentum of the initial (final) meson.
It has been shown \cite{ChengChua2003CLF,ChengChiang2010} that the
additional factor of $(-i)$ in Eq.~\eqref{eq:FFDS} follows from the demand of positive form
factors;  it can be also seen by the calculations
utilizing the heavy quark symmetry.
%For the interest of decay rate, such factor does not matter since the modulus squared of the form factors are used.
As for the $D_{(s)} \to V$ transition, we have
\begin{eqnarray}\label{eq:FFDV}
\langle V(p'',\epsilon''^{*})|V_\mu|\DDs
(p')\rangle&=&-\frac{1}{m_{\DDs}+m_V}\epsilon_{\mu\nu\alpha\beta}\epsilon''^{*\nu}p^\alpha
q^\beta V(q^2), \\
\langle V(p'',\epsilon''^{*})|A_\mu|\DDs(p')\rangle &=&
i\Biggl\{(m_{\DDs}+m_V)\left[\epsilon''^*_\mu-\frac{\epsilon''^*\cdot
p}{q^2}q_\mu\right]A_1(q^2)\nl &&-\frac{\epsilon''^*\cdot
p}{m_{\DDs}+m_V}\left[p_\mu-\frac{m_{\DDs}^2-m_V^2}{q^2}q_\mu\right]A_2(q^2)\nl
&&+2m_V\frac{\epsilon''^*\cdot p}{q^2}q_\mu A_0(q^2)\Biggr\}
\end{eqnarray}
where the relation between $A_3(q^2)$ and $A_1(q^2), A_2(q^2)$ has
been used. Finally, form factors for the $D_{(s)} \to A$ transition read
\begin{eqnarray}\label{eq:FFDA}
\langle A(p'',\epsilon''^{*})|A_\mu|\DDs
(p')\rangle&=&-\frac{1}{m_{\DDs}-m_A}\epsilon_{\mu\nu\alpha\beta}\epsilon''^{*\nu}p^\alpha
q^\beta A(q^2), \\
\langle A(p'',\epsilon''^{*})|V_\mu|\DDs(p')\rangle &=&
-i\Biggl\{(m_{\DDs}-m_A)\left[\epsilon''^*_\mu-\frac{\epsilon''^*\cdot
p}{q^2}q_\mu\right]V_1(q^2)\nl&&-\frac{\epsilon''^*\cdot
p}{m_{\DDs}-m_A}\left[p_\mu-\frac{m_{\DDs}^2-m_A^2}{q^2}q_\mu
\right]V_2(q^2)\nl && +2m_A\frac{\epsilon''^*\cdot p}{q^2}q_\mu
V_0(q^2)\Biggr\}.
\end{eqnarray}

Two remarks are in order:
\begin{itemize}
\item For the $\DDs\to A$ transition form factors, we follow the definitions
in Refs.~\cite{Cheng:2006dm,ChengChua2003CLF}, i.e., we have made
the replacements $m_{\DDs}\pm m_A \longrightarrow m_{\DDs}\mp m_A$
compared to the obsolete ones in Ref.~\cite{Cheng:2003id} since it
has been shown in \cite{ChengChua2003CLF} that such replacements will
make the transitions $B\to D_0^*,D_1$ fulfilling the similar
heavy-quark-symmetry relations as that for $B\to D, D^*$ ones.
\item In order to cancel the singularity due to
$q^2=0$, we need the constraints
\begin{eqnarray}
F^{\DDs\to S(P)}_1(0)&=&F^{\DDs\to S(P)}_0(0),\nl 2m_V
A_0(0)&=&(m_{\DDs}+m_V)A_1(0)-(m_{\DDs}-m_V)A_2(0),\nl 2m_A
V_0(0)&=&(m_{\DDs}-m_A)V_1(0)-(m_{\DDs}+m_A)V_2(0).
\end{eqnarray}
It is easily checked that the corresponding values of form factors listed in
Refs.~\cite{ChengChua2003CLF,Verma} indeed fulfill them.
\end{itemize}
As mentioned in the Introduction, Ref.~\cite{Verma} is an updated
version of Ref.~\cite{ChengChua2003CLF}, and we will stick to the
form factors obtained there based on CLFQM.

\begin{figure}
\begin{center}
\includegraphics[width=100mm]{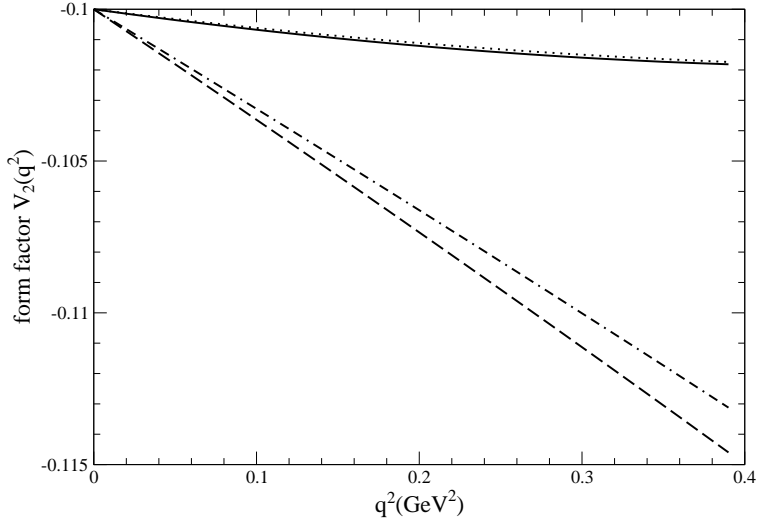}
\caption{Comparison of the different forms for the form factor
$V_2(q^2)$ corresponding to the values
$F(0)=-0.10,\,a=0.26,\,b=0.090$. The solid (dashed) line corresponds
to Eq.~\eqref{eq:Fq} (Eq.~\eqref{eq:modified Fq}), while the dotted
(dot-double-dashed) line denotes Eq.~\eqref{eq:Fq}
(Eq.~\eqref{eq:modified Fq}) with $m_D$ replaced by $m_{D_s}$.}
\label{fig:form factor}
\end{center}
\end{figure}

In CLFQM $q^+=0$ is chosen, and then $q^2=-q_\perp^2<0$, i.e., in
the spacelike region.  However, the physical situation requires
form factors be timelike ($q^2>0$). In Ref.~\cite{Jaus1990}, an
explicit form for the form factor is proposed under the assumption
that it is a continuously differentiable function of $q^2$. That
form is assumed to be valid in the full range of $q^2$, i.e., the
timelike region can be continued from the spacelike one, thus the
values in the enviroment of $q^2=0$ is crucial. The
parameters appearing in the form factor are determined by the calculation of the appropriate derivatives. In fact, the parameters can be better
determined by a fit, as has been done in
Refs.~\cite{ChengChua2003CLF,Verma}. Explicitly, we take
\begin{eqnarray}\label{eq:Fq}
F(q^2)=\frac{F(0)}{1-a (q^2/m_D^2) +b (q^2/m_D^2)^2},
\end{eqnarray}
where the values for $F(0), a, b$ corresponding to the transitions
considered in this work have been calculated in
Refs.~\cite{Verma,ChengChua2003CLF}.   As discussed in \cite{ChengChua2003CLF}, the form factor $V_2(q^2)$ for $\DDs\to A (1^{+-})$
transition approaches zero at very large $-|q^2|$ where the three-parameter parametrization (2.19) becomes questionable. To overcome this difficulty, we will fit this form factor to the form
\begin{eqnarray}\label{eq:modified Fq}
V_2(q^2)=\frac{V_2(0)}{(1-q^2/m_D^2)[1-a (q^2/m_D^2) +b
(q^2/m_D^2)^2]}.
\end{eqnarray}
In Fig.~\ref{fig:form factor}, we illustrate the difference between
Eqs.~\eqref{eq:Fq} and \eqref{eq:modified Fq} for $V_2(q^2)$ using
the same values $F(0)=-0.10,\,a=0.26,\,b=0.090$ for $V_2^{D\to b_1}$
as in Table 9 of Ref.~\cite{Verma}. Clearly, the difference between
the solid and dashed lines is large such that their integration over
$q^2$ (the area formed by the curve and $x-$axis) involved by the
differential decay rate can differ by a factor of two. One may also
consider whether $m_D$ should be replaced by $m_{D_s}$ when treating
the $D_s$ decays. In fact, the difference induced by such a
replacement is negligible, as can be seen by comparing the dotted
and solid lines or the dashed and dot-double-dashed lines in
Fig.~\ref{fig:form factor}.
%The pertinent point here we wish to express is that the modified three-parameter form, Eq.~\eqref{eq:modified Fq} is used for calculation.

In terms of the form factors given above, the
differential decay rate for $\DDs\to S (P)$  reads ($\hat m_l^2=m_l^2/q^2$)
\begin{eqnarray}
\frac{d\Gamma}{dq^2} =(1-\hat
m_l^2)^2\frac{\sqrt{\lambda(m_{\DDs}^2, m_{S(P)}^2, q^2)}G_F^2
|V_{cq}|^2}{384m_{\DDs}^3\pi^3} \Big[&&(2+\hat
m_l^2)\lambda(m_{\DDs}^2, m_{S(P)}^2, q^2) [F_1^{\DDs\to
S(P)}(q^2)]^2 \nl && +3\hat
m_l^2(m_{\DDs}^2-m_{S(P)}^2)^2[F_0^{\DDs\to S(P)}(q^2)]^2\Big],
\end{eqnarray}
with the quark flavor $q=s$ or $d$. The $\DDs\to A \ell\nu_\ell$ decay
width has the expression
\begin{align}
\frac{d\Gamma}{dq^2}&=\frac{d\Gamma_L}{dq^2}+\frac{d\Gamma_+}{dq^2}+\frac{d\Gamma_-}{dq^2},
\end{align}
with
\begin{align}
\frac{d\Gamma_L}{dq^2}&=(1-\hat
m_l^2)^2\frac{\sqrt{\lambda(m_{\DDs}^2, m_A^2, q^2)}G_F^2
|V_{cq}|^2}{384m_{\DDs}^3\pi^3}\bigg\{3\hat m_l^2\lambda(m_{\DDs}^2,
m_A^2, q^2)[V_0(q^2)]^2+\nl & \frac{2+\hat
m_l^2}{4m_A^2}\Big[(m_{\DDs}^2-m_A^2-q^2)(m_{\DDs}-m_A)V_1(q^2)-\frac{\lambda(m_{\DDs}^2,
m_A^2, q^2)}{m_{\DDs}-m_A}V_2(q^2)\Big]^2\bigg\},\label{eq:dGDA1}\\
\frac{d\Gamma_\pm}{dq^2}&=(1-\hat
m_l^2)^2\frac{\sqrt{\lambda(m_{\DDs}^2, m_A^2, q^2)}G_F^2
|V_{cq}|^2}{384m_{\DDs}^3\pi^3}\bigg\{(m_l^2+2q^2)\lambda(m_{\DDs}^2,
m_A^2, q^2)\nl &\times
\Big[\frac{A(q^2)}{m_{\DDs}-m_A}\mp\frac{(m_{\DDs}-m_A)V_1(q^2)}{\sqrt{\lambda(m_{\DDs}^2,
m_A^2, q^2)}}\Big]^2\bigg\}\label{eq:dGDA2}.
\end{align}
The $\DDs\to V \ell\nu_\ell$ decay rate can be obtained from
Eqs.~\eqref{eq:dGDA1} and \eqref{eq:dGDA2} by the following
replacements:
\begin{eqnarray}
\{A(q^2), V_0(q^2), V_1(q^2), V_2(q^2)\} &\longrightarrow& \{V(q^2),
A_0(q^2), A_1(q^2), A_2(q^2)\},\nl m_A &\longrightarrow& m_V,\nl
m_{\DDs}\mp m_A &\longrightarrow& m_{\DDs}\pm m_V.
\end{eqnarray}
Note that the form factors are real-valued. These expressions agree with
Refs.~\cite{Hong-Wei Ke,1706lattice,Wei Wang} except for some obvious
typos in Ref.~\cite{Wei Wang}.

%%%%%%%%%%%%%%%%%%%
\section{Results and discussion}
\label{Sec:results}
%%%%%%%%%%%%%%%
In this section, we consider the semileptonic decays of both $D$ and
$D_s$ mesons. For the final states, we consider the $P,\,V,\,S,\,A$
ones summarized below:
\begin{itemize}
\item $P$ is the pseudoscalar Goldstone boson $\pi, K, \eta,
\eta'$.
\item $V$ is the vector octet containing $\rho,\, \omega, \,
K^*,\,\phi$.
\item $S$ is the scalar meson lying above 1 GeV, and refers to $a_0(1450),\, f_0(1500),\,f_0(1710)$
and $K_0^*(1450)$. The state $f_0(1370)$ will not be considered by
us since its mass and width  have not been well determined
yet. PDG \cite{PDG} shows that its pole position is at $(1200-1500) -i
(150-250)$ MeV, and the Breit-Wigner or $K-$matrix mass and width
at $(1200-1500) -i (200-500)$ MeV. Note that the imaginary part of
the pole position corresponds to half of the width. From this
prospective, the main emphasis should be first concentrated on its
pole determination. Otherwise, the branching fraction cannot be
predicted in a comparable precision as other $S$ states. Concerning
the various experimental issues about the $S$ states above 1 GeV, one
may refer to the review \cite{Klempt}.
\item The axial-vector meson denoted by $A$ with the spin and parity quantum numbers
$J^P=1^-$ is classified into two categories: $1^{++}$ and $1^{+-}$.
The former contains $a_1(1260)$, $f_1(1285)$, $f_1(1420)$ while the
latter consists of $b_1(1235)$, $h_1(1170)$ and $h_1(1380)$. We will
not consider $a_1(1260)$ due to its extremely broad width
 $200-600$ MeV \cite{PDG}. We also note that $a_1(1260)$ and
$b_1(1235)$ cannot be mixed together  due to the opposite charge conjugation
parity ($C-$parity). In fact, for the fermion-antifermion pair, one
has $C=(-1)^{L+S}$ with $L$ and $S$ denoting the orbital angular
momentum and total spin between the fermion-antifermion system. But
$K_{1A}$ and $K_{1B}$ do mix together to form the physical mass
eigenstates $K_1(1270)$ and $K_1(1400)$ due to the strange and non-strange quark mass difference.
%as shown in Eq.~\eqref{eq:K1mixing} below.
\end{itemize}

The properties of the pseduscalar mesons $\pi,\,K,\,\eta$ and the
vector states $\rho,\,\omega,\,\phi,\,K^*$ are very well-known.
The $\eta$ and $\eta'$ mixing can
be written in terms of their quark states $\eta_q$ and $\eta_s$ corresponding to the $q\bar q\equiv (u\bar u
+d\bar d)/\sqrt{2}$ and $s\bar s$ components, respectively:
\begin{eqnarray}
\eta&=&\eta_q \cos\phi - \eta_s\sin\phi,\nl \eta'&=&\eta_q\sin\phi +
\eta_s\cos\phi,
\end{eqnarray}
with the mixing angle $\phi=39.3^\circ\pm1.0^\circ$
\cite{Feldmann1,Feldmann2,Feldmann3}.

We now introduce the mixing scheme for axial-vector states as well
as the scalar $f_0$ states above 1 GeV. For the axial-vector mesons,
we have \cite{ChengMixingAngle}
\begin{eqnarray} \label{eq:f1mixing}
f_1(1285)&=&f_{1q} \sin\alpha_{f1} + f_{1s} \cos\alpha_{f1},\nl
f_1(1420)&=&f_{1q}\cos\alpha_{f1}- f_{1s}\sin\alpha_{f1},
\end{eqnarray}
with $\alpha_{f1}=69.7^\circ$, and
\begin{eqnarray}\label{eq:h1mixing}
h_1(1170)&=&h_{1q} \sin\alpha_{h1}+h_{1s} \cos\alpha_{h1},\nl
h_1(1380)&=&h_{1q}\cos\alpha_{h1}-h_{1s}\sin\alpha_{h1},
\end{eqnarray}
with $\alpha_{h1}=86.7^\circ$. As before, $f_{1q}$ and $h_{1q}$
denote the $q\bar q$ component of $f_1$ and $h_1$, respectively,
while $f_{1s}$ and $h_{1s}$ denote the corresponding $s\bar s$
components. Note that in the literature, the mixing angle $\theta$
is often referred to the singlet-octet one,  and
$\alpha=\theta+54.7^\circ$ \cite{quark model PDG}. An ``ideal''
mixing is defined as $\tan \theta=1/\sqrt{2}$, i.e.,
$\theta=35.3^\circ$. Clearly, the $h_1(1170)$ is dominated by
$h_{1q}$, while the $h_1(1380)$ mainly consists of $s\bar s$. The
mixing is described at the level of state and thus the amplitude
$\mathcal A$ (or the corresponding form factor) also obeys the
relations
\begin{eqnarray}
&&\mathcal A^{D\to f_1(1285)}=\frac{1}{\sqrt{2}}\sin\alpha_{f1}
\mathcal A^{D\to f_{1q}}, \quad \mathcal{A}^{D_s\to
f_1(1285)}=\cos\alpha_{f1} \mathcal{A}^{D_s\to f_{1s}},\nl &&
\mathcal A^{D\to f_1(1420)}=\frac{1}{\sqrt{2}}\cos\alpha_{f1}
\mathcal A^{D\to f_{1q}}, \quad \mathcal{A}^{D_s\to
f_1(1420)}=-\sin\alpha_{f1} \mathcal{A}^{D_s\to f_{1s}}.
\end{eqnarray}
Note that there are no  $D\to f_{1s}$ and $D_s\to f_{1q}$
transitions. The factor of $1/\sqrt{2}$ coming from $(u\bar u+d\bar
d)/\sqrt{2}$ should be kept in mind since only the $d\bar d$
component of $f_1$ state is used in the $D\to f_1$ transition.
Similar relations also hold for $h_1$, and $f_0$ states below
\begin{eqnarray}
\mathcal A^{D\to f_0(1370)}&=&0.78/\sqrt{2}\mathcal A^{D\to
f_{0q}},\nl \mathcal A^{D\to f_0(1500)}&=&-0.54/\sqrt{2}\mathcal
A^{D\to f_{0q}},\nl \mathcal A^{D\to
f_0(1710)}&=&0.32/\sqrt{2}\mathcal A^{D\to f_{0q}},\nl
 \mathcal A^{D_s\to f_0(1370)}&=&0.51\mathcal
A^{D_s\to f_{0s}},\nl \mathcal A^{D_s\to f_0(1500)}&=&0.84\mathcal
A^{D_s\to f_{0s}},\nl\mathcal A^{D_s\to f_0(1710)}&=&0.18\mathcal
A^{D_s\to f_{0s}}.
\end{eqnarray}

The physical mass eigenstates $K_1(1270)$ and $K_1(1400)$ are the
mixture of the $^1P_1$ state $K_{1B}$ and $^3P_1$ state $K_{1A}$
\cite{quark model PDG},
\begin{eqnarray}\label{eq:K1mixing}
K_1(1270)&=&K_{1A}\sin\theta_{K_1}+K_{1B}\cos\theta_{K_1},\nl
K_1(1400)&=&K_{1A}\cos\theta_{K_1}-K_{1B}\sin\theta_{K_1},
\end{eqnarray}
and we will take $\theta_{K_1}=33^\circ$ from the analysis of
Ref.~\cite{ChengMixingAngle}. The corresponding form factors can be
obtained by
\begin{eqnarray}
\mathcal F^{\DDs\to K_1(1270)}(q^2)=\mathcal{F}^{\DDs\to
K_{1A}}(q^2)\sin\theta_{K_1}+\mathcal{F}^{\DDs\to
K_{1B}}(q^2)\cos\theta_{K_1},\nl \mathcal F^{\DDs\to
K_1(1400)}(q^2)=\mathcal{F}^{\DDs\to
K_{1A}}(q^2)\cos\theta_{K_1}-\mathcal{F}^{\DDs\to
K_{1B}}(q^2)\sin\theta_{K_1},
\end{eqnarray}
with $\mathcal{F}$ denoting $V(q^2),\, A_0(q^2),\,
A_1(q^2),\, A_2(q^2)$.

Concerning the scalar nonet with mass above 1 GeV, the $a_0(1450)$
and $K_0^*(1430)$ are believed to be the conventional $q\bar q$ mesons,
while the interpretations of $f_0(1370)$, $f_0(1500)$ and
$f_0(1710)$ are not yet achieved in full agreement, although it is
generally argued that one of them contains mainly the scalar
glueball. The controversial issue is focused on which one is primarily a
glueball. The analysis of Ref.~\cite{ChengChuaLiu} shows that the $f_0(1710)$ should have a large glueball
component and $f_0(1500)$ is mainly a flavor octet:
\begin{eqnarray}\label{eq:f0mixing}
\begin{pmatrix} f_0(1370) \\ f_0(1500) \\
f_0(1710)\end{pmatrix}=\begin{pmatrix} 0.78(2) & 0.52(3) & -0.36(1)
\\ -0.55(3) & 0.84(2) & 0.03(2) \\  0.31(1) & 0.17(1) & 0.934(4)  \end{pmatrix}
\begin{pmatrix}f_{0q} \\ f_{0s} \\G \end{pmatrix}
\end{eqnarray}
with $G$ denoting a glueball.
%In Ref.~\cite{Klempt}, $f_0(1500)$ and
%$f_0(1710)$ are understood as flavor octet states with possibly
%tetraquark composition, while the existence of $f_0(1370)$ was still
%not ascertained. The measured value of $\Gamma_{\gamma\gamma}
%\mathcal{B}(K\bar K)$ from $f_0(1710)$ decay by Belle group
%\cite{PTEP} indicates a large two-photon decay width, which
%contradicts with the expectation of much less than 1 eV  for a picture of
%glueball assignment. Thus the unmixed glueball interpretation is
%questioned by Belle. However, the mixing scheme can still not be
%excluded for our understanding.
More interesting discussions on the details can be found in
Ref.~\cite{ChengChuaLiu} in which all the existing lattice
calculations and experimental data have been considered. Hence, we adopt the mixing scheme given in Eq.~\eqref{eq:f0mixing}.
In fact, we wish to stress that the proposed measurements of
semileptonic $D$ or $D_s$ transitions to $f_0$ states will be a
powerful test for its inner structure due to the absence of the
final-state interaction between $f_0$ and the lepton pair. As least,
it can serve as a useful complement.

Based on the expressions in Sec.~\ref{Sec:FF} and the information
for the form factors in Ref.~\cite{Verma}, one can deduce the decay
rates $d\Gamma/d q^2$ for $\DDs\to M \ell^+ \nu_{\ell}$ decay and
the branching fraction  as
\begin{eqnarray}
\mathcal{B}=\frac{1}{\Gamma_{\DDs}}\int_{m_\ell^2}^{\left(m_{\DDs}-m_M\right)^2}\frac{d\Gamma}{dq^2}.
\end{eqnarray}
We refrain from repeating the values shown in Ref.~\cite{Verma}, where the
form factors for $\DDs$ decays to $P$,\,$V$,\,$S$,\,$A (1^{++})$,\,
$A (1^{+-})$ can be found in Tables 4$-$9, respectively. Our results
for the branching fractions are summarized in Table \ref{tab:BR} for
the electron decay mode and Table \ref{tab:BRmu} for the muon mode.
Strictly speaking, the $D$ decays corresponds to the charged case,
since in the CLFQM the decay constant for $D^+$ is used to determine
the $\beta$ in the vertex function
\cite{ChengChua2003CLF,Verma}.

Several remarks are in order:
\begin{itemize}
\item For $\DDs \to (P,\,V) e^+\nu_e$ decay, there are abundant experimental
data. The most recent measurements for the $D_s$ decay were done by
BESIII \cite{BESDs} and by Hietala et al. based on the CLEO
data \cite{CLEODs}. Our results are in excellent agreement with them
within errors. Especially for $D\to\pi$, $D\to
\eta$,\,$D_s\to\eta$,\,$D\to \rho$,\,$D_s\to K^*$, the central
values even match the experimental numbers exactly. Such a
surprisingly good agreement is beyond our expectation as {\it a
priori} the CLFQM does not ``know'' anything about these
experimental information.
%{\bf  --- other quantities such as quark
%masses, $\beta$ values in the wave function and decay constants are
%used as input, while the outputs are form factors and branching
%fractions.}
In other words, these values of the branching fractions can be
regarded as the predictions of CLFQM as all the input parameters
(quark masses, $\beta$ values in the vertex function) are not fitted
by the information of the measured rates. This in turn demonstrates
its predictive power. After all, we wish to stress again that an extrapolation from the space-like region to the
physical time-like one with the pole-model behavior
(Eq.~\eqref{eq:Fq}) has been utilized for form factors.
\item According to the BESIII plan \cite{BES1,BES2,Front}, $1.8\times10^7$ $D^0\bar D^0$,\,
$1.4\times10^7$ $D^+D^-$,\, $2.0\times 10^7$ $D_s^+D_s^-$ pairs will
be accumulated per year. The decays $D_s^+\to h_1(1380),\,K_1(1270)$
will be easily measured and tested. But the decays involving
$f_0(1710)$ as a final state cannot be detected currently due to the
limited statistics. However, for a super tau-charm factory, the
luminosity will be enhanced by 100 times
\cite{super1,super2,super3,super4}, and then the goal for the
measurement of these channels can be realized. We also note that
CLEO has measured the branching fraction of $D^0\to K_1^-(1270)
e^+\bar \nu_e$ with the result of $(7.6\pm4.1)\times 10^{-4}$
\cite{CLEODK1(1270)}. Considering the lifetime difference between
$D^0$ and $D^+$, our result agrees with experiment.
\item The origin for different orders of magnitudes for branching fractions
is typically understood in terms of the Cabibbo suppression and/or
phase space suppression (e.g., comparing $D\to f_0(1500)$ and $D\to
f_0(1710)$).
\item The predicted central values of the branching fractions for
$D\to \omega$,\,$D\to \bar K$,\,$D\to \bar K^*$,\,$D_s\to\phi$
semileptonic decays are in reasonable agreement with the
experimental measurements, but not as excellent as the ones for
$D\to \pi$,\,$D\to\rho$,\,$D_s\to\eta$,\,$D_s\to K^*$ as exhibited in Table
\ref{tab:BR}. In particular, the difference for the $D\to \bar K^*$ case
is a bit larger, a factor of 1.4 between theory and experiment
comparing the central values. This reminds us of some possible
theoretical errors. The main uncertainties come from form factors,
the CKM matrix elements $|V_{cs}|$ and $|V_{cd}|$, and also the
mixing angles (e.g., Eqs.~\eqref{eq:f1mixing}, \eqref{eq:h1mixing}
and \eqref{eq:f0mixing}). In Ref.~\cite{Verma}, the uncertainty of
decay constants has been propagated to the values of $\beta$ in
light-cone wave functions, otherwise, a $10\%$ variation in $\beta$
is allowed. The uncertainty arising from the form factors is
typically of the order of $2\%$. The CKM matrix elements
\begin{eqnarray}
|V_{cs}|=0.995\pm0.016,\quad |V_{cd}|=0.220\pm0.005,
\end{eqnarray}
are quoted by PDG \cite{PDG}. Considering the modulus squared, this
will yield around 5\% uncertainty. The uncertainty induced by the
mixing angle needs more care. We assign the uncertainties of
$8^\circ,\,6^\circ,\,4^\circ$ to
$\alpha_{f_1},\,\alpha_{h_1},\,\theta_{K_1}$, respectively, guided
by Ref.~\cite{ChengMixingAngle}. Allowing those variations, we get
(rough) error estimate. When the uncertainty is comparably large as
the central value, we show the resulting minimum and maximum in the
brackets. For example, the mixing angle for $h_1(1170)$ and
$h_1(1380)$ states ($\alpha_{h_1}$) can cross $90^\circ$, where the
transitions $D\to h_1(1170)$ and $D_s\to h_1(1380)$ are allowed, but
not $D_s\to h_1(1170)$ and $D\to h_1(1380)$. This shows the origin
of vanishing branching fractions of $D_s\to h_1(1170)\ell\nu_\ell$
and $D\to h_1(1380)\ell\nu_\ell$ in Tables I and II. Indeed, the
uncertainty in the mixing angles dominates the error estimate for
$D\to A$ transitions. From this point of view, it should be
understood that the BESIII measurement on these channels will be
highly meaningful for a ``precise'' determination of the mixing
angle, as also mentioned in the Introduction.

%In total, the theoretical values listed in Tables.~\ref{tab:BR} and
%~\ref{tab:BRmu} will be associated with 15\% uncertainty. Hence, we
%conclude that the branching fractions obtained from the form factors
%predicted by CLFQM agree well with experiment.
%\item The measured branching fraction for $D^+\to\bar K^{*0}e^+\nu_e$ reads
%$(5.40\pm0.10)$\% \cite{PDG} which is consistent with
%$\mathcal{B}(D^+\to\bar K^{*0}e^+\nu_e,\,\bar K^{*0}\to
%K^-\pi^+)=(3.60\pm0.07)$\% since $\mathcal{B}(\bar K^*\to
%K^-\pi^+)=2/3$ and $\mathcal{B}(\bar K^{*0}\to \bar K^0\pi^0)=1/3$.
%\item The information on the muon mode is still scarced with only two
%reports: $\mathcal{B}(D^+\to \bar K^0\mu^+\nu_\mu)=(8.74\pm0.19)$\% and
%$\mathcal{B}(D^+\to \bar K^{*0}\mu^+\nu_\mu)=(5.25\pm0.15)$\%. {\bf I'd suggest to add the experimental BFs to Table II and delete this item.}

\item We also comment on the semileptonic decay mode involving a tau lepton.
We have the masses \cite{PDG} $M(D)=1869.59\pm0.09$ MeV,
$M(D_s)=1968.28\pm0.10$ MeV, and $M(\tau)=1776.86\pm0.12$ MeV.
Hence, only the decay $D\to\tau\nu_\tau$ is allowed  by phase space,
which is constrained to be smaller than $1.2\times 10^{-3}$ with the
confidence level of 90\%. When it comes to $D_s$ decays to $\tau$
mode, $D_s^+\to\pi^0\tau^+\nu_\tau$ is also allowed besides
$D_s^+\to\tau^+\nu_\tau$. However, the aforementioned semileptonic
tau mode will be highly suppressed since there is no valence $s$
quark in the pion. One possible mechanism will be the process
$D_s^+\to\eta \tau^+\nu_\tau\to\pi^0\tau^+\nu_\tau$ via the
$\eta-\pi^0$ mixing.
\end{itemize}

We wish to comment that even-parity light mesons, including the
axial-vector meson, and the scalar meson above 1 GeV can be also
studied via $\DDs$ two-body decays \cite{Cheng:2003bn,Cheng:2010vk}
within the framework of the topological diagram approach and the
factorization scheme. The semileptonic decay modes investigated here
will provide a more clean environment to explore the nature of these
mesons owing to the absence of the strong hadronic final-state
interactions manifested in the two-body hadronic decay. Furthermore,
the size of the branching fractions considered here is of the same
order as the ones in Refs.\cite{Cheng:2003bn,Cheng:2010vk} typically
ranging from $10^{-6}$ to $10^{-3}$. So, at least, our proposal for
the semileptonic mode can be done as a supplement.
%{\bf Considering
%the known strong prediction power for the CLFQM (see the $D\to P,\,
%A$ modes in Tab.~\ref{tab:BR}), we expect the future measurements of
%the branching fractions will be consistent with our predictions
%here. Otherwise, ones need to rescrutinize the assigned mixing
%scheme, e.g., the $f_0(1500)$ is considered as a mixing state
%between the conventional $q\bar q$ structure and the glueball
%components.}

At last, we comment on the light  scalars close to or below 1 GeV,
namely the $a_0(980)$,\,$f_0(980)$ and $f_0(500)$ mesons. The
structure of these mesons are still controversial to date. One of
the popular viewpoints is to regard them  as the tetraquarks (see
e.g., Ref.~\cite{Achasov light scalar}) or the molecular states of
$\pi\pi$ and $K\bar K$ (see also a very recent work \cite{DaiLY}),
since the conventional $q\bar q$ assignment will encounter some
severe problems contradicting with experiment, see the discussions
in Refs.~\cite{Cheng:2010vk,Achasov}. A complete list of references
can be found in the reviews \cite{PDGscalar,Close}. If they are
indeed tetraquark states, it will be difficult to tackle them by the
CLFQM which is only suitable for treating the $q\bar q$ meson.
However, the attempt of probing $f_0(500)$ and $f_0(980)$ using the
CLFQM by assigning them as the $q\bar q$ configuration is available in
Ref.~\cite{Ke-f0980}, where the $q\bar q$ and $s\bar s$ mixing angle
was obtained and the $D^+\to f_0(980)e^+\nu_e$ branching fraction
was predicted. The underlying relations between the relevant form
factors are similar to those discussed above. Following the guidance
of the values presented in Ref.~\cite{ChengChua2003CLF}, the shape
parameter $\beta$ was (somewhat arbitrarily) chosen to be 0.30 allowing
10\% variation there. That is, it is not fixed by the corresponding
decay constant of the $f_0(980)$ which is zero. The vanishing decay
constants of the neutral $f_0(500),\,a_0(980),\,f_0(980)$ are the
consequence of the charge-conjugate invariance
\cite{ChengChua2003CLF}. In other words, the shape parameter cannot
be well fixed by the information of the decay constant, and instead,
other model calculations may be used.
%However, there is still
%calculation exploiting other tools, see e.g., $\DDs\to f_0(980)$
%decay in QCD sum rule \cite{Nielsen} or a very recent work for $D\to
%a_0(980)e^+\nu_e$ within the light-cone sum rule approach
%\cite{XiaodongCheng}, or just by aforementioned (quasi-) two-body
%decay process \cite{Cheng:2002ai}. {\bf Please rephrase this long
%sentence. It is hard to read.}

\begin{table}[htbp]
\centering {\scriptsize
\begin{tabular}{llllllll}
 \hline \hline
Channel &$D\to \pi$  &$D\to \bar K$  &$D\to \eta$  &$D\to\eta'$ &$D_s\to K$ &$D_s\to \eta$ &$D_s\to \eta'$\\
\hline  Theory $(10^{-2})$  &$0.41\pm0.03$ &$10.32\pm0.93$
&$0.12\pm0.01$ &$0.018\pm0.002$ &$0.27\pm0.02$
&$2.26\pm0.21$ &$0.89\pm0.09$\\
PDG ($10^{-2}$) &$0.41\pm0.02$ &$8.82\pm0.13$ &$0.11\pm0.01$
&$0.022\pm0.005$ &$0.39\pm0.09$ &$2.29\pm0.19$
&$0.74\pm0.14$\\
\hline \hline Channel &$D\to \rho$ &$D\to \omega$  &$D\to \bar K^*$
&$D_s\to K^*$ &$D_s\to \phi$ & &\\
Theory ($10^{-2}$) &$0.23\pm0.02$ &$0.21\pm0.02$ &$7.5\pm0.7$ &$0.19\pm0.02$ &$3.1\pm0.3$ & & \\
PDG ($10^{-2}$) &$0.22^{+0.02}_{-0.03}$ &$0.17\pm0.01$
&$5.40\pm0.10$
&$0.18\pm0.04$ &$2.39\pm0.23$ & & \\
\hline\hline Channel &$D\to a_0(1450)$ &$D\to f_0(1500)$ &$D\to
f_0(1710)$ &$D\to K^*_0(1450)$ &$D_s\to K^*_0(1450)$ &$D_s\to
f_0(1500)$ &$D_s\to f_0(1710)$ \\
 Theory ($10^{-5}$)  &$0.54\pm0.05$ &$0.11\pm0.02$ &$(4.7\pm0.8)\cdot10^{-4}$ &$29\pm3$ &$2.7\pm0.2$ &$15\pm3$ &$0.034\pm0.006$
 \\
 \hline\hline
 Channel &$D\to f_1(1285)$ &$D\to f_1(1420)$ &$D\to b_1(1235)$ &$D\to
 h_1(1170)$ &$D\to h_1(1380)$ &$D_s\to h_1(1170)$ &$D_s\to h_1(1380)$ \\
 Theory ($10^{-5}$) &$3.7\pm0.8$ &\{0.02, 0.14\} &$7.4\pm0.7$ &$14\pm1.5$ &\{0, 0.02\} &\{0, 19.7\} &$64\pm7$ \\ \hline\hline
 Channel &$D\to K_1(1270)$ &$D\to K_1(1400)$ &$D_s\to K_1(1270)$  &$D_s\to
 K_1(1400)$ &$D_s\to f_1(1285)$ & $D_s\to f_1(1420)$& \\
 Theory ($10^{-5}$) &$320\pm40$ &\{0.5, 2.0\} &$17\pm2$ &\{0.05, 0.14\} &\{6.0, 36\} &$25\pm5$ &\\
 \hline \hline
\end{tabular}}
 \caption{Braching fractions of $D^+$ and $D_s^+$ decays to $(P,\,V,\,S,\,A) e^+ \nu_e$. Units are shown in
 parentheses. PDG average values are taken from Ref.~\cite{PDG}, while data are not yet available for the
 $S$ and $A$ modes. When the error bar is comparable to the central value, instead we show the
 minimum and maximum values in the brackets.}
 \label{tab:BR}
\end{table}

\begin{table}[htbp]
\centering {\scriptsize
\begin{tabular}{llllllll}
 \hline \hline
Channel &$D\to \pi$  &$D\to \bar K$  &$D\to \eta$  &$D\to\eta'$ &$D_s\to K$ &$D_s\to \eta$ &$D_s\to \eta'$\\
Theory $(10^{-2})$  &$0.41\pm0.03$ &$10.07\pm 0.91$ &$0.12\pm0.01$
&$0.017\pm0.002$ &$0.26\pm0.02$
&$2.22\pm0.20$ &$0.85\pm0.08$\\
PDG ($10^{-2}$) & & $8.74\pm0.19$ &  & & & & \\
 \hline \hline
Channel &$D\to \rho$ &$D\to \omega$  &$D\to \bar K^*$
&$D_s\to K^*$ &$D_s\to \phi$ & &\\
Theory ($10^{-2}$) &$0.22\pm0.02$ &$0.20\pm0.02$ &$7.0\pm0.7$ &$0.19\pm0.02$ &$2.9\pm0.3$ & & \\
PDG ($10^{-2}$) & & & $5.25\pm0.15$ & & & &\\
 \hline\hline Channel &$D\to a_0(1450)$ &$D\to
f_0(1500)$ &$D\to f_0(1710)$ &$D\to K^*_0(1450)$ &$D_s\to
K^*_0(1450)$ &$D_s\to
f_0(1500)$ &$D_s\to f_0(1710)$ \\
 Theory ($10^{-5}$)  &$0.38\pm0.03$ &$0.07\pm0.01$ &$(2.5\pm0.4)\cdot10^{-5}$ &$22\pm2.0$ &$2.2\pm0.2$ &$12\pm2$ &$0.014\pm0.002$\\
 \hline\hline
 Channel &$D\to f_1(1285)$ &$D\to f_1(1420)$ &$D\to b_1(1235)$ &$D\to
 h_1(1170)$ &$D\to h_1(1380)$ &$D_s\to h_1(1170)$ &$D_s\to h_1(1380)$ \\
 Theory ($10^{-5}$) &$3.2\pm0.6$ &\{0.02, 0.12\} & $6.4\pm0.6$ &$12.2\pm1.3$ &\{0, 0.02\} &\{0, 17.4\} &$54\pm6$ \\ \hline\hline
 Channel &$D\to K_1(1270)$ &$D\to K_1(1400)$ &$D_s\to K_1(1270)$  &$D_s\to
 K_1(1400)$ &$D_s\to f_1(1285)$ & $D_s\to f_1(1420)$ & \\
 Theory ($10^{-5}$) &$260\pm30$ &\{0.4, 1.7\} &$15\pm2$ &\{0.05, 0.12\} &\{5.2, 30.6\} &$21\pm5$ &\\
 \hline \hline
\end{tabular}}
 \caption{Same as Table \ref{tab:BR} but for the muon mode, i.e.,
 $D^+$ and $D_s^+$ decays to $(P,\,V,\,S,\,A) \mu^+ \nu_\mu$. }
 \label{tab:BRmu}
\end{table}

\section{Conclusion}
The covariant light-front model is a powerful tool to predict
the electroweak decay form factors. In Ref.~\cite{ChengChua2003CLF},
the authors have systematically calculated the form factors for $D$
transition  to $S$- and $P$-wave mesons. The extension to the
$D_s^+$ decay has been done in Ref.~\cite{Verma}, where the
parameters $\beta$ in the light-front wave functions were
constrained by the available experimental information as well as the
lattice results. Based on the form factors there, we have calculated
the branching fractions for various channels of the $D$ and $D_s$
decays to $(P,\,S,\,V,\,A)\, \ell\bar\nu_\ell$, with $P,\,S,\,V,\,A$
denoting a pseudoscalar, scalar above 1 GeV, vector and
axial-vector, respectively, and $\ell=e$ or $\mu$. Results are shown
in Table~\ref{tab:BR} for the electron decay mode and Table
\ref{tab:BRmu} for the muon mode. Comparing to the available
experimental data, we find the covariant light-front model works
very well. The branching fractions for other channels are also
predicted. The semileptonic decay mode provides a clean environment
to examine the hadron structures. The experimental searches are
pointed out. Most of them can be measured by the BESIII
Collaboration, while for a future super tau-charm factory, the
statistics will be enhanced by 100 times. These future measurements
confronting with the theoretical predictions here will definitely
shed light on our basic understanding of the semileptonic $D$ and
$D_s$ decay as well as the inner structure of the relevant scalars
with masses above 1 GeV and axial-vector mesons. Other approaches
for probing the structures of the scalar and axial-vectors are
compared and commented.

{\bf Acknowledgments:} XWK is grateful to  Hong-Wei Ke, Wei Wang,
Hsiang-Nan Li for useful discussions  and Hai-Bo Li for the
insightful discussions on the BES measurments. He also specially
acknowledges R.~C.~Verma for the clarification of Ref.~\cite{Verma}.
This work is supported by the Ministry of Science and Technology of
R.O.C. under Grant No. 104-2112-M-001-022.

%%%%%%%%%%%%%%%%

\end{document}